\documentclass[a4paper,11pt]{article}
\usepackage{amsfonts}
\usepackage{amsmath}

\input{tcilatex}

\begin{document}

\author{Duoje Jia$^{\thanks{%
E-mail: jiadj@nwnu.edu.cn; Tel: 086-0931-3385702; Fax:086-0931-7971503 \ }}$
\\
Institute of Theoretical Physics, College of Physics\\
and Electronic Engineering, Northwest Normal \\
University,\textit{\ Lanzhou 730070, P.R. China}\\
Interdisciplinary Center for Theoretical Study, \\
University of Science \& Technology of China, \\
Hefei, Aanhui 230026\textit{, P.R. China}}
\title{An Effective Dual Abelian-Higgs Model from SU(2) Yang-Mills theory
via Connection Decomposition}
\maketitle

\begin{abstract}
It has long argued that confinement in non-Abelian gauge theories, such as
QCD, can be account for by analogy with\ typed II superconductivity. In this
paper, we show that it is possible to arrive at an effective dual
Abelian-Higgs model, the dual and relativistic version of Ginzburg-Landau
model for superconductor, from $SU(2)$ Yang-Mills theory based on the
Faddeev-Niemi connection decomposition and the order-disorder assumptions
for the gauge field. The implication of these assumptions is discussed and
role of the resulted scalar field is analyzed associated with the
"electric-magnetic" duality and theory vacuum. It is shown that the mass
generation of the gauge vector field can arise from quantum fluctuation of
the coset basis variable $\partial \mathbf{n}$, and the mass of the
"electric" field is approximately equal to that of the scalar particle. A
generalized dual London equation with topologically quantized singular
vortices is derived for the static "electric" field from the our dual model.

PACS number(s): 11.15Tk, 12,38.Aw, 12.38.Aw

\end{abstract}

\section{Introduction}

Three decades ago, the dual-superconductor(DS) picture of confinement in
quantum chromodynamics (QCD) was proposed by Nambu, Mandelstam and 't Hooft
\cite{Nambu,tHooft78}, in an analogy to\ type II superconductivity. In DS
picture, color and quarks are confined within hadrons due to the dual
Meissner effect. This picture was further elaborated by idea of Abelian
projection \cite{tHooftB455} that QCD can be reduced, by partially-fixing
gauge to Maximal Abelian(MA) gauge, to an Abelian gauge theory with magnetic
charges forming condensate. This idea was confirmed by lattice calculations
(see \cite{Greensite03,Suzuki,Suganuma} for a review) for long-distance
gluondynamics.

On the other hands, an proposal was put forward by 't Hooft\cite{tHooftA03}
that an gauge-invariants scalar kernel $Z(\phi )$ should exist in the
effective model for confinement as relicts of the infrared counter term.
However, the its implication in framework of Yang-Mills(YM) theory remains
unknown, even if the color confinement is believed to take place in the pure
gluondynamics\cite{Shifman}. By deriving an effective Abelian-Higgs(AH)
model from SU(2) YM theory based on a series of assumptions, one argument
\cite{Dzhunushaliev03} suggests that $\phi $ in AH model can arise from
off-diagonal gauge bosons. The assumptions include the order-stochastic
assumptions for the diagonal and the off-diagonal parts of gauge vector
field, the de-correlation assumption between the two parts and simple-mode
approximation for the correlation function of off-diagonal field. However,
in such an approach, the connections of the scalar to the magnetic charges
and "electronic-magnetic" duality was not seen clearly.

The purpose of this paper is to point out that it is possible to derive an
effective dual AH model starting from $SU(2)$ YM theory, based on the three
assumptions that is similar to the ideas in Ref. \cite{Dzhunushaliev03}. In
this model, the scalar $\phi $ comes from off-diagonal gauge degrees of
freedom and the gauge-invariants scalar kernel $Z(\phi )$ appears as an
effective magnetic-media factor for theory vacuum, whose the vacuum
expectation value(vev.) can provide a scalar potential of Mexico-hat form.
This is done by reformulating YM theory via the connection decomposition (CD)%
\cite{Duan,Cho80,FN}, and then use the order-disorder assumptions for the
new variables in CD. We also use\ the de-correlation assumptions between the
dual variables and three local basis in iso-space. The mass generation of
the Abelian field can arise from quantum fluctuation of the off-diagonal
basis $\partial \mathbf{n}$, and is approximately equal to that of the
scalar particle. A generalized dual London equation with topologically
quantized singular vortices was derived for the static "electric" field from
the dual AH\ model.

\section{Abelian projection in terms of connection decomposition}

\bigskip In the approach of CD\cite{Duan, Cho80, FN}, the separating the
infrared variables from YM connection can be done by introducing infrared
unit order-parameter field $\mathbf{n}(x)$. Based on CD, the vacuum
structure of the YM theory and gluonball spectrum are studied, associated
with knotted-vortex excitations\cite{Faddeev,Cho}. Here, we study DS picture
in YM\ theory from the viewpoint of CD approach.

We consider $SU(2)$ YM theory where connection("gluon" field) $\mathbf{A}%
_{\mu }=A_{\mu }^{a}\tau ^{a}$ ($\tau ^{a}=\sigma ^{a}/2,a=1,2,3$) describes
6 transverse ultraviolet degrees of freedom. To parameterize $\mathbf{A}%
_{\mu }$ in terms of new variables, we invoke an unit iso-vector $\mathbf{n}$%
. Solving $\mathbf{A}_{\mu }$ from $D_{\mu }\mathbf{n}-\partial _{\mu }%
\mathbf{n}=g\mathbf{A}_{\mu }\times \mathbf{n}$, where $g$ is coupling
constant, one gets\cite{Duan}
\begin{equation}
\mathbf{A}_{\mu }=A_{\mu }\mathbf{n}+g^{-1}\partial _{\mu }\mathbf{n}\times
\mathbf{n+b}_{\mu }  \label{deD}
\end{equation}%
where $A_{\mu }\equiv \mathbf{A}_{\mu }\cdot \mathbf{n}$ transforms as an
Abelian connection for $U(1)$ rotation $U(\alpha )=e^{i\alpha n^{a}\tau ^{a}%
\text{ }}$round the iso-direction $\mathbf{n}$ ($A_{\mu }\rightarrow A_{\mu
}+\partial _{\mu }\alpha /g$) and $\mathbf{b}_{\mu }=g^{-1}\mathbf{n}\times
D_{\mu }(\mathbf{A}_{\mu })\mathbf{n}$ is $SU(2)$ covariant. Here, the
Abelian part $A_{\mu }\mathbf{n}$ in (\ref{deD}) corresponds to Abelian
subgroup $H=U(1)$, while the non-Abelian gluon parts $\mathbf{C}_{\mu
}=g^{-1}\partial _{\mu }\mathbf{n}\times \mathbf{n}$ and $\mathbf{b}_{\mu
}=b_{\mu }^{a}\tau ^{a}$, both of which are orthogonal to $\mathbf{n}$,
correspond to coset group $SU(2)/H$. We note that (\ref{deD}) can be true
variable change \cite{Shabanov} if one takes $\mathbf{b}_{\mu }$ itself as a
gauge vector field and further imposes two constraints on\textbf{\ }$\mathbf{%
b}_{\mu }$. This is necessary for getting marginal contribution to the final
effective action but we do not consider such a contribution in this paper.

The fact that $\mathbf{C}_{\mu }$ does not depend upon the original degrees
of freedom $\mathbf{A}_{\mu }$ implies $\mathbf{A}_{\mu }$ has intrinsic
structure. This idea is firstly due to the work on multi-monopoles\cite{Duan}
and has been generalized to the $SO(N)$-connection case\cite{Lee,MXH} as
well as the spinorial-decomposition case\cite{Jinstant,Fu}. Further
decomposing $\mathbf{b}_{\mu }$ in terms of the local basis \{$\partial
_{\mu }\mathbf{n}$, $\partial _{\mu }\mathbf{n}\times \mathbf{n}$\} of the
internal coset space $SU(2)/H$, one find CD \cite{FN} for $SU(2)$ connection

\begin{equation}
\mathbf{A}_{\mu }=A_{\mu }\mathbf{n}+\mathbf{C}_{\mu }+g^{-1}\phi
_{1}\partial _{\mu }\mathbf{n}+g^{-1}\phi _{2}\partial _{\mu }\mathbf{n}%
\times \mathbf{n,}  \label{deF}
\end{equation}%
in which $A_{\mu }$ (and $\mathbf{A}_{\mu }$) has dimension of mass, $%
\mathbf{n}$, the scalars $\phi _{1}$ and $\phi _{2}$ of unit.

The transformation role of the new variables in (\ref{deF}) under the gauge
rotation $U(\alpha )$ can be found by requiring the CD\ (\ref{deF}) to be
covariant under $U(\alpha )$. Clearly, $\mathbf{C}_{\mu }$ is intact under $%
U(\alpha )$. The transformation role of $\phi _{1}$ and $\phi _{2}$ can be
given by the $U(\alpha )$-covariance of $\mathbf{b}_{\mu }$. In fact, one
finds
\begin{eqnarray*}
(b_{\mu }^{a}\tau ^{a})^{U} &=&g^{-1}e^{i\alpha \mathbf{n}\cdot \mathbf{\tau
}}(\phi _{1}\partial _{\mu }n^{a}\tau ^{a}+\phi _{2}\epsilon ^{abc}\partial
_{\mu }n^{b}n^{c}\tau ^{a}\mathbf{)}e^{-i\alpha \mathbf{n}\cdot \mathbf{\tau
}}, \\
\ &=&g^{-1}(\phi _{1}-\alpha \phi _{2})\partial _{\mu }n^{a}\tau
^{a}+g^{-1}(\phi _{2}+\alpha \phi _{1})(\partial _{\mu }\mathbf{n}\times
\mathbf{n})^{a}\tau ^{a},
\end{eqnarray*}%
which implies $\delta \phi _{1}=-\alpha \phi _{2}\ $and $\delta \phi
_{2}=\alpha \phi _{1}$, or $\delta (\phi _{1}+i\phi _{2})=i\alpha (\phi
_{1}+i\phi _{2})$. Thus, the complex variables $\phi =\phi _{1}+i\phi _{2}$
transforms as a charged complex scalar:%
\begin{equation*}
\phi \rightarrow \phi e^{i\alpha }.
\end{equation*}

In comparison with $12$ field components of $\mathbf{A}_{\mu }$, the new
variables ($A_{\mu }$,$n^{a}$,$\phi $) in (\ref{deF}) has $8(=4+2+2)$
degrees of freedom, making CD\ (\ref{deF}) a singular variable change. The
CD then defines a singular gauge-fixing in which $A_{\mu }\mathbf{n}$ is the
unfixed diagonal part of the vector field $\mathbf{A}_{\mu }$ and the other
terms in RHS of (\ref{deF}) are non-Abelian(off-diagonal) components. The
singularities in CD\ (\ref{deF}) are caused by the difference between two
group manifolds $SU(2)$ and $H$, and is determined by the singularities in $%
\mathbf{n}(x)$ \cite{Shabanov}. The singularities in $\mathbf{n}(x)$ is due
to the global parameterization of a gauge group with nontrivial topology in
terms of local basis, and they responds to magnetic monopoles configuration,
as shown below.

In fact, if one chooses $\mathbf{n}$ as hedgehog configuration\ $\mathbf{n}=%
\mathbf{x}/r$ then $\mathbf{C}_{\mu }$ becomes the Wu-Yang potential \cite%
{Wu-Yang} for non-Abelian magnetic monopole:$A_{i}^{a}=\epsilon
_{aic}x^{c}/r^{2}$. The field strength $\mathbf{C}_{\mu \nu }=\partial _{\mu
}\mathbf{C}_{\nu }-\partial _{\nu }\mathbf{C}_{\mu }$ for $\mathbf{C}_{\mu }$
can be written as $\mathbf{C}_{\mu \nu }\equiv B_{\mu \nu }\mathbf{n}$ where
\begin{equation}
B_{\mu \nu }\equiv -g^{-1}\mathbf{n}\cdot (\partial _{\mu }\mathbf{n}\times
\partial _{\nu }\mathbf{n})  \label{h}
\end{equation}%
stands for the magnetic field strength. One can identify the (projected)
magnetic potential $C_{\mu }$ by definition $B_{\mu \nu }\equiv \partial
_{\mu }C_{\nu }-\partial _{\nu }C_{\mu }$, where the parametrization of $%
C_{\mu }$ can not be given globally. This can be seen by parameterizing $%
\mathbf{n}$ in terms of spherical coordinates as $\mathbf{n}=(\sin \gamma
\cos \beta ,\sin \gamma \sin \beta ,\cos \gamma )$. One get from (\ref{h})
\begin{eqnarray*}
B_{\mu \nu } &=&-g^{-1}\sin \gamma (\partial _{\mu }\gamma \partial _{\nu
}\beta -\partial _{\nu }\gamma \partial _{\mu }\beta ) \\
C_{\mu } &=&g^{-1}(\cos \gamma \partial _{\mu }\beta \pm \partial _{\mu
}\alpha ),
\end{eqnarray*}%
which takes the form of the $SU(2)$ magnetic configuration. $C_{\mu }$ is
not uniquely defined and has a gauge freedom of $U(1)$ transformation ($%
C_{\mu }\rightarrow C_{\mu }+g^{-1}\partial _{\mu }\alpha ^{\prime }$),
corresponding to the rotation $U(\alpha ^{\prime })$ round $\mathbf{n}$.
This $U(1)$ covariance is happened to hold simultaneously for Abelian part $%
A_{\mu }$. By setting $\mathbf{n}$ along $\sigma ^{3}$, one sees that
choosing a local direction $\mathbf{n}(x)$ at each point $x$ ensures the
covariance of CD for $U(\alpha ^{\prime })$-rotation, but not for the
rotation round $\sigma ^{1}$ or $\sigma ^{2}$. The singularities of $\mathbf{%
n}(x)$ occurs at points where the orientation of $\mathbf{n}(x)$ is not
defined due to the global reparametrization of the field variables in terms
of the local basis ($\mathbf{n}$, $d\mathbf{n,}d\mathbf{n\times }d\mathbf{n}%
) $.

By comparison CD (\ref{deF}) with the global decomposition%
\begin{equation}
\mathbf{A}_{\mu }=A_{\mu }^{3}\tau ^{3}+A_{\mu }^{1}\tau ^{1}+A_{\mu
}^{2}\tau ^{2},  \label{uv}
\end{equation}%
one finds that there are the local correspondences between $A_{\mu }\mathbf{n%
}$ and diagonal part $A_{\mu }^{3}\tau ^{3}$, and ($\phi _{1}$, $\phi _{2})$%
-terms and ($A_{\mu }^{1}\tau ^{1}$, $A_{\mu }^{2}\tau ^{2})$. However, no
correspondence exists in (\ref{uv}) for the potential $\mathbf{C}_{\mu }$.
If we parameterize the SU(2) matrix in terms of Euler angles ($\alpha
(x),\beta (x),\gamma (x)$):
\begin{equation*}
U(x)=e^{i\beta \sigma ^{3}/2}e^{i\gamma \sigma ^{2}/2}e^{i\alpha \sigma
^{3}/2}=\left(
\begin{array}{cc}
e^{i(\beta +\alpha )/2}\cos (\gamma /2) & -e^{i(\beta -\alpha )/2}\sin
(\gamma /2) \\
e^{-i(\beta -\alpha )/2}\sin (\gamma /2) & e^{-i(\beta +\alpha )/2}\cos
(\gamma /2)%
\end{array}%
\right)
\end{equation*}%
one finds
\begin{eqnarray}
C_{\mu }(x) &=&tr\left( \sigma ^{3}\frac{i}{g}U(x)\partial _{\mu }U^{\dag
}(x)\right)  \label{n1} \\
n^{a}(x) &=&2Tr\left( \sigma ^{a}U^{\dag }(x)\tau ^{3}U(x)\right)  \label{n2}
\end{eqnarray}%
i.e., the magnetic potential $C_{\mu }$ originates from the pure gauge $%
iU\partial _{\mu }U^{\dag }/g$. Therefore, when we choose $U(x)$ to be the
singular transformation mapping the global basis \{$\tau ^{1\sim 3}$\} to
the local basis \{$\mathbf{n},\partial _{\mu }\mathbf{n},\partial _{\mu }%
\mathbf{n}\times \mathbf{n}$\} and set $\mathbf{n}(\infty )=\mathbf{n}_{0}$%
(constant unit vector) we get one example of Abelian projection, where the
maximal Abelian subgroup $H$ responds to the gauge rotation round $\mathbf{n}
$. This implies that CD makes the basis dynamic and contains the magnetic
variable as a topological degree of freedom. The monopole occurs at the
singularities of the transformation $U(x)$ or iso-vector $\mathbf{n}(x)$, as
can be seen from (\ref{n1}) and (\ref{n2}).

With (\ref{deF}), one finds the gauge field strength:
\begin{eqnarray}
\mathbf{G}_{\mu \nu } &=&\mathbf{n[}F_{\mu \nu }+Z(\phi )B_{\mu \nu }]+g^{-1}%
\left[ \nabla ^{\mu }\phi \mathbf{n}_{\nu }-\nabla _{\nu }\phi \mathbf{n}%
_{\mu }\right]  \notag \\
&&+\frac{1}{2g}\left[ \nabla _{\mu }\phi \mathbf{n}_{\nu }-\nabla _{\nu
}\phi \mathbf{n}_{\mu }\right]  \label{G}
\end{eqnarray}%
where $F_{\mu \nu }\equiv \partial _{\mu }A_{v}-\partial _{v}A_{\mu }$, $%
Z(\phi )=1-|\phi |^{2}$ and $\nabla _{\mu }\phi \equiv \nabla _{\mu }\phi
_{1}+i\nabla _{\mu }\phi _{2}=(\partial _{\mu }-igA_{\mu })\phi $ is $U(1)$
covariant derivative where $\nabla _{\mu }\phi _{1}\equiv \partial _{\mu
}\phi _{1}+gA_{\mu }\phi _{2}$ and $\nabla _{\mu }\phi _{2}\equiv \partial
_{\mu }\phi _{2}-gA_{\mu }\phi _{1}$. Here, $\mathbf{n}_{\mu }=\partial
_{\mu }\mathbf{n}-i\partial _{\mu }\mathbf{n}\times \mathbf{n}$ and $h.c.$
stands for Hermite conjugation.

It is suggestive to consider the Abelian components of the full gauge field $%
\mathbf{G}_{\mu \nu }$
\begin{eqnarray}
\mathbf{G}_{\mu \nu } &\rightarrow &\mathbf{n}[F_{\mu \nu }+H_{\mu \nu }]%
\text{,}  \label{abelian} \\
H_{\mu \nu } &:&=Z(\phi )B_{\mu \nu }.  \notag
\end{eqnarray}%
This gives a hint for "electric-magnetic" duality ($A_{\mu }\leftrightarrow
C_{\mu }$, $F_{\mu \nu }\leftrightarrow H_{\mu \nu }$ ) in an effective
"magnetic media" described by the off-diagonal variable $\phi $. Similar
media-like factor $Z(\phi )$ is introduced phenomenologically by 't Hooft%
\cite{tHooftA03} to account for the QCD vacuum. The Eq.(\ref{abelian}) can
also be reached by taking large-$g$ limit, which transforms the standard YM
theory into an pure Abelian gauge theory with "electric" charge as well as
"magnetic" charge. One can see from (\ref{abelian}) that owing to the
off-diagonal field the original "electric-magnetic" duality given by 't
Hooft tensor $f_{\mu \nu }=F_{\mu \nu }+B_{\mu \nu }$ becomes that in
effective "magnetic media".

\section{Order-disordered assumption for gauge connection}

\bigskip To approach the infrared YM\ theory non-perturbatively, we will use
three assumptions about the gauge field variables, which is similar to, but
not same with that presented in Ref. \cite{Dzhunushaliev03}. We argued that
these assumptions is valid approximately in far-infrared regime. The three
main assumptions are listed as below:

(1). The off-diagonal basis $\partial _{\mu }\mathbf{n}$ in CD becomes
stochastic while $\mathbf{n}$ is ordered in quantum YM\ theory. That is, one
has following assumption on the vev:
\begin{eqnarray}
\langle \partial _{\mu }n^{a}(x)\rangle &=&0,\text{ and }\langle \partial
_{\mu }n^{a}(x)\partial ^{\nu }n^{b}(x)\rangle \neq 0,  \label{ass1} \\
\text{only if }\mu &=&\nu ,a=b  \notag
\end{eqnarray}%
and $\langle \partial _{\mu }n^{a}n^{b}\rangle =0$ even if $a=b$. This is
consistent with the gauge, Lorentz and parity invariance of the ground
state. The nonvanishing vev. is looked as a consequence of the quantum
fluctuation of $\partial _{\mu }\mathbf{n}$. Eq.(\ref{ass1}) means $\langle
\partial ^{\mu }n^{a}\partial _{\nu }n^{a}\rangle \propto \delta _{\nu
}^{\mu }$.

(2). The Abelian field $A_{\mu }$ behaves as classical variable, and the
dual variables ($A_{\mu },\phi $) are de-correlated with basis $\mathbf{n}$
and $\partial _{\mu }\mathbf{n}$. That means
\begin{equation}
\langle F(A_{\mu },\phi )P(\mathbf{n}\text{,}\partial \mathbf{n})\rangle
=\langle F(A_{\mu },\phi )\rangle \langle P(\mathbf{n}\text{,}\partial
\mathbf{n})\rangle ,  \label{ass2}
\end{equation}%
where $F$ and $P$ are any functions of the involved variables. The classical
behavior of $A_{\mu }$ means that $\langle f(A_{\mu })\rangle \approx
f(A_{\mu })$ being a good approximation for any function $f$ of $A_{\mu }$.
This has close analogy to the semi-classical treatment of the radiation
field, where the radiation field behaves as classical variable while the
matter degrees of freedom are taken to be the quantum operators.

(3). After quantization of theory, the complex variable $\phi ^{\ast }(x)$
is taken to be a field operator $\phi ^{\dag }$ of creating a charged scalar
particle at $x$, and $\phi (x)$ to be the the corresponding operator
annihilating that the particle. In the vacuum state, $\phi $ shows
off-diagonal long range order (ODLRO), which can be explicitly written as
\begin{equation}
\langle \phi (x)\phi ^{\dag }(y)\rangle =\Phi (x)\Phi ^{\ast }(y),\text{for }%
x_{0}>y_{0}.  \label{ass3}
\end{equation}

One may doubt the justification of above three assumptions. It is true, to
my knowledge, that no direct evidence for these assumptions exists. However,
the indication that the off-diagonal gluon amplitude is strongly suppressed
and off-diagonal gluon phase shows strong randomness is indeed observed in
lattice simulation \cite{Suganuma03,Suganuma0407}. Furthermore, the
order-disorder transition\cite{Giacomo,Carmona-DEliaD0203} is also predicted
numerically in which an operator of creating magnetically-charged particle
similar to $\phi ^{\dag }$ is shown to be the order parameter for this
transition. The same results was also seen in the numerical analysis of the
specific heat and of the chiral order parameter at the chiral transition\cite%
{05}. We note that the ODLRO assumption for $\phi $ in (3) does not
contradict with the randomness of the off-diagonal gauge potential in the
sense that the later is caused by the randomness of the local basis field \{$%
\partial _{\mu }\mathbf{n}$, $\partial _{\mu }\mathbf{n}\times \mathbf{n}$\}
of the coset group $SU(2)/H$. One can ready verify by Eqs. (\ref{ass1}) and (%
\ref{ass2}) that $\mathbf{b}_{\mu }$ is stochastic:$\langle \mathbf{b}_{\mu
}\rangle =0$ but $\langle (\mathbf{b}_{\mu })^{2}\rangle \neq 0$. In
addition, the classical behavior of $A_{\mu }$ was always assumed in DS
analysis (see \cite{Giacomo97} and references therein) of the infrared QCD
and was predicted in the lattice simulations\cite%
{Giacomo,Suganuma03,Suganuma0407}.

The most subtle ansatz may be the assumption (\ref{ass2}). It implies that
the "magnetic" field does not couple with AH variables ($A_{\mu }$,$\phi $)
at the classical level. When combining with the ansatz (\ref{ass1}), this
implies the variables $\mathbf{n}$ in (\ref{deF}) forms the background field
in with contrast ($A_{\mu }$,$\phi $), as we will shown in the next section.
We will see there that the ansatz (\ref{ass2}) is necessary for us to derive
the dual AH model and the combining of three assumptions will lead to the
Bogomolnyi limit\cite{Bogomolnyi} $m_{\Phi }\approx m_{A}$, being consistent
with numerical result \cite{Gubarev99}.

\subsection{Effective dual Abelian-Higgs action}

\bigskip\ \ \ Putting (\ref{G}) into the YM Lagrangian $\mathfrak{L}=-%
\mathbf{G}_{\mu \nu }^{2}/4$, one gets%
\begin{equation}
\mathfrak{L}_{dual}=-\frac{1}{4}F_{\mu \nu }^{2}+\mathfrak{L}_{V}+\mathfrak{L%
}_{FB}+\mathfrak{L}_{D},  \label{dual}
\end{equation}%
where
\begin{equation*}
\mathfrak{L}_{V}=-\frac{1}{4}Z(\phi )^{2}B_{\mu \nu }^{2}
\end{equation*}%
\begin{equation*}
\mathfrak{L}_{FB}=-\frac{1}{2}Z(\phi )F_{\mu \nu }B^{\mu \nu }
\end{equation*}%
\begin{equation*}
\mathfrak{L}_{D}=-\frac{1}{2g^{2}}(n_{\mu \nu }-igB_{\mu \nu })(\nabla ^{\mu
}\phi )^{\dag }\nabla ^{\nu }\phi
\end{equation*}%
Here, $n_{\mu \nu }\equiv \eta _{\mu \nu }(\partial _{\rho }\mathbf{n}%
)^{2}-\partial _{\mu }\mathbf{n}\cdot \partial _{\nu }\mathbf{n}$ and $\phi $
appears as a charged field minimally coupled with 'electric' field $A_{\mu }$
by strength $g$.

\bigskip We look all variables in (\ref{deF}) as quantum operators. Since
the ground state is both the (relativistic and gauge) rotation and
translation invariant, the stochastic assumption (1) for $\partial \mathbf{n}
$ implies
\begin{equation}
\ \langle (\partial _{x}\mathbf{n})^{2}\rangle =\langle (\partial _{y}%
\mathbf{n})^{2}\rangle =\langle (\partial _{z}\mathbf{n})^{2}\rangle
=(\partial _{0}\mathbf{n})^{2}=m^{2}  \label{pns}
\end{equation}%
\begin{equation}
\langle (\partial _{\mu }n^{a})^{2}\rangle =\frac{1}{3}\langle (\partial
_{\mu }\mathbf{n})^{2}\rangle =-\frac{2}{3}m^{2}\text{,for fixed }a=1,2,3.
\label{pnf}
\end{equation}%
in which $m$ is a mass scale of the $\partial \mathbf{n}$\textbf{-}%
fluctuation. This gives%
\begin{equation}
\langle (\partial \mathbf{n})^{2}\rangle :=\langle \partial ^{\mu
}n^{a}(x)\partial _{\mu }n^{a}(x)\rangle =-2m^{2}.  \label{pn}
\end{equation}%
From (\ref{ass1}),\ (\ref{pns}) and (\ref{pn}), one has%
\begin{equation}
\left\langle n_{\nu }^{\mu }\right\rangle =\delta _{\nu }^{\mu }\left\langle
(\partial \mathbf{n})^{2}\right\rangle -\delta _{\nu }^{\mu }\left\langle
(\partial _{0}\mathbf{n})^{2}\right\rangle =-3\delta _{\nu }^{\mu }m^{2}
\label{nmn}
\end{equation}

Using Wick theorem and the assumption (1), one can show
\begin{eqnarray*}
\langle B_{\mu \nu }\rangle &=&g^{-1}\epsilon ^{abc}\underset{\varepsilon
^{\mu }\rightarrow 0}{\lim }\partial _{\mu }^{x_{2}}\partial _{\nu
}^{x_{3}}\langle n^{a}(x_{3})n^{b}(x_{2})n^{c}(x_{1})\rangle \\
&=&g^{-1}\epsilon ^{abc}\ \underset{x_{3}\rightarrow x}{\lim }\ \langle
n^{a}(x_{3})\rangle \langle \partial _{\mu }n^{b}(x_{2})\partial _{\nu
}n^{c}(x_{1})\rangle |_{x_{2}=x_{1}=x} \\
&\propto &\epsilon ^{abc}\langle \partial _{\mu }n^{b}(x)\partial _{\nu
}n^{c}(x)\rangle =0,
\end{eqnarray*}%
where $\varepsilon ^{\mu }$ $\equiv max_{i,j}\left\Vert (x_{i})^{\mu
}-(x_{j})^{\mu }\right\Vert $ is four small positive parameters ($i,j=1,2,3$%
) and the time-order was assumed so that $%
(x)^{0}<(x_{1})^{0}<(x_{2})^{0}<(x_{3}\,)^{0}<(x)_{0}+\varepsilon ^{0}$
before taking limit.

Including the full off-diagonal variables and applying the assumptions (1)$%
\sim $(3) to the reformulated YM Lagrangian (\ref{dual}), one can calculate
the vacuum average of the Lagrangians as below. First, one has%
\begin{eqnarray}
\langle \mathfrak{L}_{V}\rangle &=&-(\lambda /4)\,\left\langle 1+(\phi
^{\dag }\phi )^{2}-2\phi ^{\dag }\phi \right\rangle  \notag \\
&=&-(\lambda /4)\left( 1+2|\Phi ^{\ast }\Phi |^{2}-2\Phi ^{\ast }\Phi \right)
\notag \\
&=&-(\lambda /2)\left[ (|\Phi |^{2}-1/2)^{2}+1/4\right]  \label{Lv}
\end{eqnarray}%
where $\lambda \equiv \langle B_{\mu \nu }^{2}\rangle $ is positive scale
and with dimension of 4. We have used the assumption (2) and (3), the Bose
symmetry of the scalar field and the Wick theorem:%
\begin{eqnarray*}
\langle \phi ^{\dag }\phi \phi ^{\dag }\phi \rangle |_{x} &=&\underset{%
\varepsilon ^{\mu }\rightarrow 0}{\lim }\langle \phi _{1}^{\dag }\phi
_{_{2}}\phi _{3}^{\dag }\phi _{4}\rangle \\
&=&\underset{\varepsilon ^{\mu }\rightarrow 0}{\lim }\{\langle \phi _{4}\phi
_{3}^{\dag }\rangle \langle \phi _{2}\phi _{1}^{\dag }\rangle +\langle \phi
_{4}\phi _{1}^{\dag }\rangle \langle \phi _{3}\phi _{2}^{\dag }\rangle
+\langle \phi _{1}^{\dag }\phi _{3}^{\dag }\rangle \langle \phi _{2}\phi
_{4}\rangle \} \\
&=&2\Phi (x)\Phi ^{\ast }(x)\Phi (x)\Phi ^{\ast }(x).
\end{eqnarray*}%
Here, $\phi _{i}\equiv \phi (x_{i})$ and $\varepsilon ^{\mu }$ are the
maximal norm of $(x_{i})^{\mu }-(x_{j})^{\mu }$, where $i,j=1\sim 4$. The
time-order is assumed so that $(x)^{0}<(x_{1})^{0}<\cdots
<(x_{4}\,)^{0}<(x)_{0}+\varepsilon ^{0}$. Moreover, $\langle \phi _{1}^{\dag
}\phi _{3}^{\dag }\rangle =\langle \phi _{2}\phi _{4}\rangle =0$ for $%
x_{1}=x_{3}$, $x_{2}=x_{4}\,$since in the Fock state of vacuum only paired
products of $\phi $ and $\phi ^{\dag }$ has nonvanishing vev.. Furthermore,
\begin{eqnarray*}
\langle \lbrack \nabla ^{\mu }\phi (x)]^{\dag }\nabla _{\mu }\phi (x)\rangle
&=&\langle (\partial _{\mu }\phi ^{\dag }(x)\partial ^{\mu }\phi
(x)+igA^{\mu }(x)\phi ^{\dag }(x)\partial _{\mu }\phi (x) \\
&&-igA_{\mu }(x)\partial ^{\mu }\phi ^{\dag }(x)\phi (x)+g^{2}A^{\mu
}(x)A_{\mu }(x)\phi ^{\dag }(x)\phi (x)\rangle \\
&=&\underset{y\rightarrow x^{-}}{\lim }[\partial _{x}^{\mu }\partial _{\mu
y}+igA^{\mu }\partial _{\mu x}-igA_{\mu }\partial ^{\mu y}+g^{2}A^{\mu
}A_{\mu }]\left\langle \phi (x)\phi ^{\dag }(y)\right\rangle ,
\end{eqnarray*}%
which leads to%
\begin{eqnarray}
\langle \lbrack \nabla ^{\mu }\phi ]^{\dag }\nabla _{\mu }\phi \rangle
&=&\partial ^{\mu }\Phi ^{\ast }(x)\partial _{\mu }\Phi (x)+igA^{\mu
}(x)\partial _{\mu }\Phi (x)\Phi ^{\ast }(x)  \notag \\
&&-igA_{\mu }(x)\Phi (x)\partial ^{\mu }\Phi ^{\ast }(x)+g^{2}A^{\mu
}(x)A_{\mu }(x)\Phi (x)\Phi ^{\ast }(x)  \notag \\
&=&[\nabla _{\mu }\Phi (x)]^{\ast }\nabla ^{\mu }\Phi (x)  \label{par}
\end{eqnarray}%
Combining with (\ref{nmn}) and the relation $\langle B_{\mu \nu }(\nabla
^{\mu }\phi )^{\dag }\nabla ^{\nu }\phi \rangle =0$, this gives
\begin{equation}
\langle \mathfrak{L}_{D}\rangle =\frac{3m^{2}}{2g^{2}}|(\partial _{\mu
}-igA_{\mu })\Phi |^{2}  \label{Ld}
\end{equation}

Noticing that $\left\langle \mathfrak{L}_{FB}\right\rangle \propto \langle
B^{\mu \nu }\rangle =0$ due to the assumptions (2), and putting (\ref{Lv}), (%
\ref{Ld}) into (\ref{dual}), one gets
\begin{eqnarray}
\mathfrak{L}^{eff} &=&-\frac{1}{4}F_{\mu \nu }^{2}+\frac{3m^{2}}{g^{2}}%
|\nabla ^{\mu }\Phi |^{2}  \notag \\
&&-\frac{\lambda }{2}(|\Phi |^{2}-\frac{1}{2})^{2}-\frac{\lambda }{8}
\label{L1}
\end{eqnarray}%
\ Since $\lambda >0$ has dimension of 4 and is proportional to $g^{-2}$ one
can rewrite it as $\lambda =m^{\ast 4}/g^{2}$, in which $m^{\ast 4}$ is a
positive mass scale determined by:%
\begin{equation}
m^{\ast 4}=\left\langle (\mathbf{n},d\mathbf{n}\times d\mathbf{n}%
)^{2}\right\rangle .  \label{mst}
\end{equation}

By scaling $\Phi $ into that with dimension of mass%
\begin{equation}
\sqrt{\frac{3}{2}}\frac{m}{g}\Phi (x)\rightarrow \Phi (x),  \label{rep}
\end{equation}%
and ignoring an additive constant in (\ref{L1}), we arrive at the effective
dual AH Lagrangian
\begin{equation}
\mathfrak{L}^{eff}=-\frac{1}{4}F_{\mu \nu }^{2}+|(\partial _{\mu }-igA_{\mu
})\Phi |^{2}-V(\Phi )  \label{AHM}
\end{equation}%
where the potential $V(\Phi )$ of the scalar $\Phi $ has the Mexico-hat form
\begin{eqnarray}
V(\Phi ) &=&\frac{\widetilde{\lambda }}{4}(|\Phi |^{2}-\mu ^{2})^{2}.
\label{v} \\
\widetilde{\lambda } &=&\frac{8g^{2}}{9}(\frac{m^{\ast }}{m})^{4},\mu =\frac{%
\sqrt{3}m}{2g}>0.  \notag
\end{eqnarray}%
\ \ \ \

We can determine $m^{\ast }$ in (\ref{mst}) by using the assumption (1).
Noticing the relation (\ref{ass1}) and using Wick theorem, we find
\begin{eqnarray*}
m^{\ast 4} &=&\epsilon _{abc}\epsilon _{mkl}[\langle \partial _{\mu
}n^{b}\partial ^{\mu }n^{k}\partial _{\nu }n^{c}\partial ^{\nu
}n^{l}n^{a}n^{m}\rangle ] \\
&=&\epsilon _{abc}\epsilon _{mkl}[\langle \partial _{\mu }n^{b}\partial
^{\mu }n^{k}\rangle \langle \partial _{\nu }n^{c}\partial ^{\nu
}n^{l}\rangle \langle n^{a}n^{m}\rangle +\langle \partial _{\mu
}n^{b}\partial ^{\nu }n^{l}\rangle \langle \partial _{\nu }n^{c}\partial
^{\mu }n^{k}\rangle \langle n^{a}n^{m}\rangle \\
&&+\langle \partial _{\mu }n^{b}n^{m}\rangle \langle \partial _{\nu
}n^{c}\partial ^{\nu }n^{l}\rangle \langle \partial ^{\mu }n^{k}n^{a}\rangle
+\cdots ] \\
&=&\epsilon _{abc}\epsilon _{mkl}\delta ^{bk}\delta ^{cl}\langle (\partial
_{\mu }n^{1})^{2}\rangle ^{2}\langle n^{a}n^{m}\rangle \\
&=&2!(\frac{-2m^{2}}{3})^{2}\delta ^{am}\langle n^{a}n^{m}\rangle \\
&=&\frac{8}{9}m^{4}.
\end{eqnarray*}%
where we have used the relation (\ref{pnf}). With (\ref{pns}), one has $%
\left\langle (\nabla \mathbf{n})^{2}\right\rangle =3m^{2}$. Thus
\begin{eqnarray}
\widetilde{\lambda } &=&(\frac{8g}{9})^{2}  \notag \\
m &=&\left\langle (\nabla \mathbf{n})^{2}\right\rangle ^{1/2}/\sqrt{3}
\label{m1}
\end{eqnarray}%
It follows that the mass of the complex scalar $\Phi $ is
\begin{equation*}
m_{\Phi }=\sqrt{\widetilde{\lambda }}\mu =\frac{4\sqrt{3}}{9}m,
\end{equation*}

The model (\ref{AHM}) is well known as the dual AH model in the original
dual-superconductor mechanism\cite{Nambu,tHooft76} for the confining phase
of QCD, as an analogy to the Ginzburg-Landou model for superconductor. The
dual Higgs mechanism for the model (\ref{AHM}) will enable $A_{\mu }$ to
acquires a mass.

The Hamiltonian with respect to the Lagrangian (\ref{AHM}) is
\begin{eqnarray*}
\mathcal{H} &=&\int d^{3}x\left[ |D^{0}\Phi |^{2}+|\vec{D}\Phi |^{2}+\frac{1%
}{2}(\partial _{t}\vec{A}+\nabla A^{0})^{2}\right. \\
&&\left. +\frac{1}{2}\vec{B}^{2}+V(\Phi )\right] ,
\end{eqnarray*}%
which has a lowest energy solution $A_{\mu }(x)=0,\Phi (x)=\mu ,(\mu >0)$.
Here, $\vec{B}:=\nabla \times \vec{A}$ and the global phase factor has
canceled for simplicity. Making the shift $\Phi (x)=\mu +\eta (x)$ enables
us to rewrite Lagrangian as
\begin{equation}
\mathfrak{L}^{eff}=-\frac{1}{4}F_{\mu \nu }^{2}+\frac{3m^{2}}{4}A_{\mu
}A^{\mu }+(\partial _{\mu }\eta )^{2}-V(\eta ),  \label{Lsb}
\end{equation}%
where $\eta $ can be taken to be a real scalar by choosing unitary gauge,
and it has potential
\begin{equation}
V(\eta )=\frac{16g^{2}}{81}\eta ^{2}[\eta +2\mu ]^{2}.  \label{vsb}
\end{equation}

Clearly, $V(\eta )$ favors the zero vev. of $\eta $. Thus, the Higgs
mechanism leads $A_{\mu }$ to acquires a mass
\begin{equation}
m_{A}=\sqrt{\frac{3}{2}}m  \label{mA}
\end{equation}%
and the complex scalar $\Phi $ becomes into a real scalar particle $\eta $
with mass%
\begin{equation}
m_{\eta }=m_{\Phi }=\frac{4\sqrt{3}}{9}m,  \label{mass}
\end{equation}%
as can be seen in (\ref{Lsb}) and (\ref{vsb}). Then, one has%
\begin{equation*}
\frac{m_{A}}{m_{\Phi }}\approx \frac{3}{2}
\end{equation*}

We see here that, as a consequences of order-disorder and de-correlation
transition of the variables ($A_{\mu },\mathbf{n},\phi $) in infrared limit,
the effective theory of YM$\ $theory becomes dual AH model, in which both
the Abelian "electric" field $A_{\mu }$ and the scalar $\Phi $(and $\eta $)
acquire the masses $\sim m$.\ The mass scale $m$ arises from the the quantum
fluctuation $\left\langle (\nabla \mathbf{n})^{2}\right\rangle ^{1/2}$ of
the stochastic field $\nabla \mathbf{n}$ and has the same order with $%
m_{\eta }\ $and $m_{A}$, though the determination of this mass scale quite
nontrivial. This is in agreement with the the lattice simulation\cite%
{Gubarev99} where the Bogomolnyi limit\cite{Bogomolnyi} $m_{\Phi }\approx
m_{A}$ for AH model is predicted by fitting the SU(2) Lattice gauge theory%
\cite{Schilling98}, as is observed in many Lattice QCD calculation\cite%
{Kato98}.

One can also see that the effective potential $V(\Phi )$ which favors the
SSB of the AH model (\ref{AHM}) is developed from the averaged media-factor $%
\langle Z(\phi )\rangle $ and the resulted masses ($m_{\eta },m_{A}$) for
scalar and "electric" field do not depend explicitly on the coupling $g$,
though the form of potential $V(\eta )$ does. This favors the idea of DS
picture that the YM vacuum is of the "magnetic" type in far-infrared limit%
\cite{Nambu,tHooft76}. A similar proof\cite{Niemi05} is also proposed
recently that the SU(2) YM theory can reduce to a two-band dual
superconductor with an interband Josephson coupling.

\subsection{Generalized Dual London equation}

\bigskip There are much evidences that vortices can be responsible for color
confinement:Vortex configurations reproduce a great number of the asymptotic
force between a static quark-antiquark pair (see \cite{Greensite03} for a
review). As a key ingredient in DS picture of confinement, such a
topological configuration was recently shown \cite{Gattnar04} to be
necessary in the light of Gribov-Zwanziger's confinement condition.

Here, we show that our effective AH\ model (\ref{AHM}) can give rise to a
dual generalized London's equation for Abelian "electric" field $A_{\mu }$,
which allows the Abrikosov-Nielsen-Olesen(ANO) vortex-string of the
"electric" field at the zero of $\Phi $. The equation of motion of the model
(\ref{AHM}) is
\begin{eqnarray}
\partial _{\nu }F^{\mu \nu } &=&-ig(\Phi ^{\ast }\partial ^{\mu }\Phi
-\partial ^{\mu }\Phi ^{\ast }\Phi )-2g^{2}|\Phi |^{2}A^{\mu }  \notag \\
\nabla _{\mu }\nabla ^{\mu }\Phi &=&-\frac{\widetilde{\lambda }}{2}[|\Phi
|^{2}-\mu ^{2}]\Phi \text{, }  \label{EL}
\end{eqnarray}%
where the Lorenz gauge ($\partial _{\mu }A^{\mu }=0$) is used.

It is suggestive to first consider the uniform solution of $\Phi $ to Eq.(%
\ref{EL}). It is
\begin{eqnarray}
\Phi &\approx &\mu \text{,}  \notag \\
\partial _{\nu }F^{\mu \nu } &=&j^{\mu }=-m_{A}^{2}A^{\mu }.  \label{London}
\end{eqnarray}

We note that the second equation can also be obtained by taking the large-$g$
limit in (\ref{EL}), where $\Phi (x)\sim g^{-1}$, as seen in (\ref{rep}).
The second equation of Eq.(\ref{London}) is known as London's relation,
which is valid only in the uniform or slow-varying situation of $\Phi (x)$.
The parameter $m_{A}$, given by (\ref{mA}), is the mass scale responsible
for dual Meissner effect, and its inverse $\lambda _{L}=1/m_{A}\sim m^{-1}$
determines the transverse dimensions of the "electric" field $A_{\mu }$
penetrating into the YM\ vacuum. This length scale is about $0.95$fm for the
data $m_{A}=1.3$ Gev given in Ref \cite{Gubarev99}. For finite $g$, the
numerical solutions for "electric" field and $\Phi $ are studied in Ref.
\cite{Schilling98,Gubarev99}. Below, we presents pure topological argument
that the ANO\ vortex must exist at the classical level as "electric" vortex
line by taking the zero of $\Phi $ into account.

We consider the static case of Eq.(\ref{EL}). Instead of writing $\Phi
=|\Phi |e^{iS(x)}$ which enables $\vec{V}:=-i\Phi ^{\ast }%
\overleftrightarrow{\nabla }\Phi /(2|\Phi |^{2})$ to take the form of
velocity potential ($\vec{V}\propto \nabla S(x)$), we write $\Phi $ as $\Phi
=\Phi _{1}+i\Phi _{2}$, with real scalars $\Phi _{1}$ and $\Phi _{2}$
forming a vector field $\vec{\Phi}:=(\Phi ^{1},\Phi ^{2})$. Denote its
direction field as $N^{i}=$ $\Phi ^{i}/|\Phi |,(i=1.2)$ and rewrite $\vec{V}$
as $\vec{V}=\epsilon _{ij}N^{i}\nabla N^{j}$. Then, the vorticity of $\vec{V}
$ becomes
\begin{eqnarray*}
(\nabla \times \vec{V})^{a} &=&\epsilon ^{abc}\epsilon _{ij}\partial
_{b}[N^{i}\partial _{c}N^{j}] \\
&=&\epsilon ^{abc}\epsilon _{ij}\partial _{b}[\frac{\Phi ^{i}}{|\Phi |^{2}}%
]\partial _{c}\Phi ^{j} \\
&=&\epsilon ^{abc}\epsilon _{ij}[\frac{\partial ^{2}}{\partial \Phi
^{i}\partial \Phi ^{k}}\ln |\Phi |]\partial _{b}\Phi ^{k}\partial _{c}\Phi
^{j} \\
&=&\triangle _{\Phi }\ln |\Phi |J^{a}(\Phi /x).
\end{eqnarray*}

If we define a vectorial Jacobian $J^{a}(\Phi /x)$ of $\Phi $ and use the
Laplace relation in $\Phi $-space(here, $\triangle _{\Phi }:=$ $\partial
^{2}/\partial \Phi ^{i}\partial \Phi ^{i}$ ):
\begin{eqnarray}
\epsilon ^{ij}J^{a}(\Phi /x) &:&=\epsilon ^{abc}\partial _{b}\Phi
^{i}\partial _{c}\Phi ^{j},  \label{J} \\
\triangle _{\Phi }\ln |\Phi | &=&2\pi \delta ^{2}(\vec{\Phi}),  \notag
\end{eqnarray}%
then, one gets%
\begin{equation}
(\nabla \times \vec{V})^{a}=2\pi \delta ^{2}(\vec{\Phi})J^{a}(\Phi /x).
\label{curlV}
\end{equation}

From the first equation of (\ref{EL}), one has
\begin{equation}
\vec{A}+\lambda (x)^{2}\nabla \times \vec{B}=\frac{1}{g}\vec{V}  \label{ABV}
\end{equation}%
where $\lambda (x)=1/(\sqrt{2}g|\Phi |)$ is called effective penetrating
length here in analogy to the typed-I superconductor. Taking curl of
equation (\ref{ABV}) yields%
\begin{equation}
\vec{B}-\lambda ^{2}\nabla ^{2}\vec{B}-4g^{2}\lambda ^{4}(\vec{\Phi}\partial
_{j}\vec{\Phi})[\nabla B^{j}-\partial ^{j}\vec{B}]=\frac{\mathcal{\phi }_{0}%
}{2\pi }\nabla \times \vec{V}.  \label{BFV}
\end{equation}%
where $\mathcal{\phi }_{0}=2\pi /g$ is the unit quanta of vortex flux, as
shown below.

We show that the vorticity in (\ref{curlV}) can be written as the sum of
delta-functions of the lines by using the formula\cite{MXH}%
\begin{eqnarray}
\delta ^{2}(\vec{\Phi}) &=&\sum_{k}w_{k}(\vec{\Phi})\int_{L_{k}}ds\frac{%
\delta ^{3}(\vec{r}-\vec{r}_{k}(s))}{D(\frac{\Phi }{u})_{\Sigma _{k}}}
\label{delF} \\
J(\frac{\Phi }{u})_{\Sigma _{k}} &=&\det \left[ \frac{\partial \Phi ^{m}}{%
\partial u^{l}}\right] ,(i,m=1,2)  \label{Du}
\end{eqnarray}%
where $L_{k}\,\ $stands for the zero-lines $\overrightarrow{r}_{k}(s)$ of $%
\Phi $ given by $\Phi ^{1}(x,y,z)=0$ ,$\Phi ^{2}(x,y,z)=0$, $w_{k}(\vec{\Phi}%
)$ is the winding number of map $\vec{\Phi}:x\rightarrow \Phi (x)$ around
the singular line $L_{k}\,$, and $\Sigma _{k}$ stands for the transverse
section of $L_{k}$ at $\vec{r}_{k}(s)$, with ($u^{1},u^{2}$) being the
surface parameters. One can show from (\ref{J}) and (\ref{Du}) that
\begin{equation*}
\left. \frac{\vec{J}(\frac{\Phi }{x})}{J(\frac{\Phi }{u})_{\Sigma _{k}}}%
\right\vert _{\overrightarrow{r}_{k}(s)}=\frac{d\vec{r}_{k}}{ds}
\end{equation*}%
which, combining with (\ref{delF}) and (\ref{curlV}), leads to
\begin{equation*}
\nabla \times \vec{V}=2\pi \sum_{k}w_{k}(\vec{\Phi})\int_{L_{k}}d\vec{r}%
_{k}\delta ^{3}(\vec{r}-\vec{r}_{k}).
\end{equation*}%
Integration of $\nabla \times \vec{V}$ over surface $\Sigma =\cup \Sigma
_{k} $, the collection of all transverse section of $L_{k}$, gives the
circulation quantization of $\vec{V}$: $\doint $ $\vec{V}\cdot d\vec{l}=2\pi
n$, where $n=\sum_{k}w_{k}$ is a topological integer. Then, Eq.(\ref{BFV})
becomes
\begin{equation}
\vec{B}-\lambda ^{2}\nabla ^{2}\vec{B}-4g^{2}\lambda ^{4}(\vec{\Phi}\partial
_{j}\vec{\Phi})[\nabla B^{j}-\partial ^{j}\vec{B}]=\mathcal{\phi }%
_{0}\sum_{k}w_{k}(\vec{\Phi})\int_{L_{k}}d\vec{r}_{k}\delta ^{3}(\vec{r}-%
\vec{r}_{k}),  \label{GLD}
\end{equation}%
which is the generalized London's equation. The RHS of Eq.(\ref{GLD})
represents the quantized ANO\ vortices, with unit flux quanta of $\mathcal{%
\phi }_{0}=2\pi /g$. In the situation that $\Phi (x)$ varys very slowly over
the space (analogy to London limit for typed II superconductor), Eq.(\ref%
{GLD}) becomes
\begin{equation*}
\vec{B}-\lambda ^{2}\nabla ^{2}\vec{B}=\mathcal{\phi }_{0}\sum_{k}w_{k}(\vec{%
\Phi})\int_{L_{k}}d\vec{r}_{k}\delta ^{3}(\vec{r}-\vec{r}_{k}).
\end{equation*}%
which is the generalization of the London's equation $\vec{B}-\lambda
^{2}\nabla ^{2}\vec{B}=0$ to vortex-state case\cite{ZhangB}. Eq.(\ref{GLD})
shows that the quantized flux for each vortex is given by $\mathcal{\phi }%
_{0}w_{k}(\vec{\Phi})$, the integer multiple of $\mathcal{\phi }_{0}$.

It should be pointed out that the RHS of Eq.(\ref{GLD}) includes all
topological contributions due to the vortices and is valid for many
vortex-line situation. One can see from the above deduction that this
topological term does not depend upon the dynamical details of the scalar $%
\Phi (x)$ and is due to the topological quantization of the flux of $\vec{B}$%
.

\subsection{Concluding remark}

We have shown that the effective dual AH action can follows from $SU(2)$ YM
theory, based on the Faddeev-Niemi decomposition of $SU(2)$ gauge field and
several assumptions. The analysis of the implication of these assumptions
and the role of the resulted scalar field in the effective model suggests
that one may reasonably look field $\Phi $ as the field of monopole and
anti-monopole, whose superfluidity induces dual "repulsive effects" with
respect to the "electric" field $A_{\mu }$. This agrees with the analysis
\cite{Giacomo} of the magnetic charge operator. The mass generation for the
"electric" field as well as the scalar can be due to quantum fluctuation of
the local coset basis $\partial \mathbf{n}$. The approximate equality of
masses between the "electric" field and the scalar is observed, being in
agreement with the prediction of Ref.\cite{Gubarev99}. We also derive a
generalized dual London equation with topologically quantized singular
string for the static "electric" field from the dual AH model (\ref{AHM}).
We hope this will shed a light on the studies of the DS picture for
color-confinement mechanism within the framework of YM theory.

In the opinion of this paper, the effective theory of YM theory in the
far-infrared limit is that of the AH variable ($A_{\mu }$,$\phi $) at the
classical level, with $\mathbf{n}$ being the quantum background. This
differs with the opinion of Ref. \cite{Faddeev} that the effective theory of
infrared YM theory is that of $\mathbf{n}$ variable and that the theory
vacuum is described by knotted string. However, in the absence of the
external sources, it is favorable for vortex line in Eq.(\ref{GLD}) to form
closed or knotted string due to the tension of vortex tube. The analysis
here implies that the vortex tube seems to be the "electric" knot, as was
argued in \cite{Cho}.

We also note that the analysis of this paper is based on the on-shell CD (%
\ref{deF}), which ignores the marginal terms in the effective model (\ref%
{AHM}). The inclusion of such a term can be done by appying the approach in
this paper to the off-shell CD \cite{Faddeev,JDJ}. The details of the
relevant analysis will be presented in elsewhere.

\textbf{Acknowledgements}

D. Jia thanks X-J Wang and J-X Lu for numerous discussion, and M-L. Yan for
valuable suggestions. This work is supported by the Postdoctoral Fellow
Startup Fund of NWNU (No. 5002--537).

\end{document}